\documentclass[12pt]{iopart}
\usepackage{times}
\usepackage{mathptmx}
\usepackage{graphicx}
\usepackage[dvips,colorlinks]{hyperref}

\begin{document}

\centerline { \bf LINEAR SECOND-ORDER DIFFERENTIAL EQUATIONS} 
\centerline{\bf }
\centerline{\bf FOR BAROTROPIC FRW COSMOLOGIES}

\bigskip 

\centerline {H.C. Rosu$^1$\footnote{hcr@ipicyt.edu.mx}, O. Cornejo$^1$, M. Reyes$^2$, D. Jimenez$^2$} 

\bigskip

\centerline{$^1$ Dept. of Appl. Math., IPICyT, Apdo Postal 3-74 Tangamanga, San Luis Potos\'{\i}, MEXICO} 

\centerline{$^2$ Instituto de F\'{\i}sica, Universidad de Guanajuato, Apdo Postal E-143, Le\'on, MEXICO}

\bigskip



\bigskip

\noindent
{
{\bf Abstract} 

\noindent
Simple linear second-order differential equations have been written down for FRW cosmologies with
barotropic fluids by Faraoni. His results have been extended by Rosu, who employed techniques belonging to nonrelativistic supersymmetry to obtain
time-dependent adiabatic indices.
Further extensions are presented here using the known connection between the linear second-order differential equations and Dirac-like equations in the same
supersymmetric context. These extensions are equivalent to adding an imaginary part to the adiabatic index which is proportional to
the mass parameter of the Dirac spinor. The natural physical interpretation of the imaginary part is related to the particular dissipation and 
instabilities of the barotropic FRW hydrodynamics that are introduced by means of this supersymmetric scheme.  \\}   
 
\bigskip

\noindent

\bigskip


{\bf 1} - {\bf Introduction}

\noindent
The barotropic FRW cosmologies obey the following set of differential equations:

\begin{equation} \label{e1}
{\rm \frac{\ddot{a}}{a}=-\frac{4\pi G}{3}(\rho +3p)}~,
\end{equation}
\begin{equation} \label{e2}
{\rm
\left(\frac{\dot{a}}{a}\right)^2=\frac{8\pi G\rho}{3}-\frac{\kappa}{a^2}~,}
\end{equation}
\begin{equation} \label{e3}
{\rm p=(\gamma -1)\rho,}
\end{equation}
where ${\rm a}$ is the scale factor of the universe, $\rho$ and ${\rm p}$ are the energy
density and the pressure, respectively,
of the perfect fluid of which a classical universe is usually assumed to be
made of, $\kappa=0,\pm1$ is the curvature index of the flat, closed, open
universe, respectively, and $\gamma$ is the constant adiabatic index of the cosmological fluid.
Recently, Faraoni \cite{Far} proposed  the ``Riccati route" of solving the system of eqs. (1)-(3)
and Rosu used Faraoni's approach to introduce a supersymmetric class of cosmological fluids
possessing time-dependent adiabatic indices \cite{rosu}. 
It was claimed that these fluids can provide a simple
explanation for a currently accelerating universe \cite{type1a}. 

In this work, we review the supersymmetric factorization methods for barotropic FRW cosmologies in section 2.
Next, in section 3, we present corresponding Dirac-like (first-order) coupled differential equations and their associated 
second-order differential equations and discuss them in a formal way. We end up the work with a short conclusion
section.

\newpage

\bigskip
{\bf 2} - {\bf Supersymmetric (factorization) methods}

\noindent
Combining the equations (1)-(3) and using the conformal time variable $\eta$  defined by
${\rm dt=a(\eta)d\eta}$ one gets the equation
\begin{equation} \label{comb}
{\rm \frac{a^{''}}{a}+(c-1)\left(\frac{a^{'}}{a}\right)^2+c\kappa=0~.}
\end{equation}
where ${\rm c=\frac{3}{2}\gamma -1}$. The case $\kappa=0$ is directly
integrable \cite{Far} and will be skipped henceforth.
One can immediately see that by means of the change of function
${\rm u=\frac{a^{'}}{a}}$ the following Riccati equation is obtained
\begin{equation} \label{ricc}
{\rm u^{'}+cu^2+\kappa c=0~.}
\end{equation}
Employing now ${\rm u=\frac{1}{c}\frac{w^{'}_{\kappa}}{w_{\kappa}}}$ one gets the
very simple second order differential equation
\begin{equation} \label{schr1}
{\rm w^{''}_{\kappa}+\kappa c^2w_{\kappa}=0~.}
\end{equation}
For $\kappa =1$ the solution of the latter is
${\rm w_{1}=W_1\cos(c\eta +d)}$, where ${\rm d}$ is an arbitrary phase,
implying
$
{\rm a_{1}(\eta)=A_1[\cos(c\eta +d)]^{1/c}~,}
$
whereas for $\kappa =-1$ one gets
${\rm w_{-1}=W_{-1}{\rm sinh}(c\eta)}$ and therefore
$
{\rm a_{-1}(\eta)=A_{-1}[{\rm sinh}(c\eta)]^{1/c}}~,
$
where ${\rm W_{\pm 1}}$ and ${\rm A_{\pm 1}}$ are amplitude parameters.
These are the same solutions as in the textbook procedures.


The point now is that the Riccati solution
${\rm u_{p}=\frac{1}{c}\frac{w^{'}}{w}}$
mentioned above is only the particular solution, i.e.,
${\rm u_{p,1}=-\tan (c\eta )}$ and ${\rm u_{p,-1}={\rm coth} (c\eta )}$
for $\kappa =\pm 1$, respectively. The particular Riccati solutions are
closely related to the common factorizations of the second-order linear differentail
equations that are directly related to the well-known Darboux isospectral transformations
\cite{ms}.
Indeed, Eq.~(\ref{schr1}) can be written
\begin{equation} \label{w}
{\rm w^{''}-c(-\kappa c)w=0}
\end{equation}
and also in factorized form using Eq.~(6) one gets (${\rm D_{\eta}=\frac{d}{d\eta}}$)
\begin{equation} \label{w3}
{\rm \left(D_{\eta}+cu_{p}\right)
\left(D_{\eta}-cu_{p}\right)w=
w^{''}-c(u_{p}^{'}+cu_{p}^{2})w=0}~.
\end{equation}
To fix the ideas, we shall call Eq.~(\ref{w3}) the bosonic equation.
On the other hand, the supersymmetric partner (or fermionic)
equation of Eq.~(\ref{w3}) will be
\begin{equation} \label{f}
{\rm
\left(D_{\eta}-cu_{p}\right)
\left(D_{\eta}+cu_{p}\right)w_f=
w^{''}_{f}-c(-u_{p}^{'}+cu_{p}^2)w_{f}={\rm w^{''}_f
-c\cdot c_{\kappa, susy}w_f=0}}~.
\end{equation}
Thus, one can write
$$
{\rm c_{\kappa,susy}(\eta)=-u_{p}^{'}+cu_{p}^2=
\left\{ \begin{array}{ll}
{\rm c(1+2{\rm tan}^2 c\eta)} & \mbox{if $\kappa =1$}\\
{\rm c(-1+2{\rm coth}^2 c\eta)} & \mbox{if $\kappa =-1$}
\end{array} \right.}
$$
for the supersymmetric partner adiabatic index.
The solutions $\rm w_f$ are $\rm w_f =\frac{c}{\cos (c\eta +d)}$ 
and $\rm w_f =\frac{c}{sinh (c\eta)}$ for $\kappa =1$ and $\kappa =-1$,
respectively. 

Introducing the (quantum momentum) operator ${\rm P_{\eta}}=-i{\rm D_{\eta}}$ we can write the fermionic 
equations as follows
\begin{equation} \label{Pf}
{\rm
\left(-P_{\eta}-icu_{p}\right)
\left(P_{\eta}-icu_{p}\right)w_f=
-P_{\eta}^{2}w_{f}-c(-iP_{\eta}u_{p}+cu_{p}^2)w_{f}
}
~,
\end{equation}
whereas the bosonic case is
\begin{equation} \label{Pb}
{\rm
\left(P_{\eta}-icu_{p}\right)
\left(-P_{\eta}-icu_{p}\right)w_b=
-P_{\eta}^{2}w_{b}-c(iP_{\eta}u_{p}+cu_{p}^2)w_{b}
}
~,
\end{equation}

There is a more general factorization of the bosonic equation \cite{M84}
 \begin{equation} \label{wbg}
{\rm \left(D_{\eta}+cu_{g}\right)
\left(D_{\eta}-cu_{g}\right)w_g=
w_g^{''}-c(u_{g}^{'}+cu_{g}^{2})w_g=w_g^{''}+\kappa cc(\eta ; \lambda)w_g=0}~,
\end{equation}
which is given in terms of the general Riccati solution  ${\rm u_{g}(\eta)}$ 
\begin{equation} \label{ug}
{\rm u_{g}}(\eta;\lambda)=  {\rm u_p(\eta)} -\frac{1}{{\rm c}} {\rm D_{\eta}}
\Big[ {\rm ln}({\rm I_{\kappa}}(\eta) + \lambda) \Big]
={\rm  D_{\eta}
\Big[ ln \left(\frac{w_{\kappa}(\eta)}{{\rm I_{\kappa}}(\eta) +
\lambda}\right)^{\frac{1}{c}}\Big]}
\end{equation}
and yields the one-parameter family of
adiabatic indices ${\rm c_{\kappa}(\eta;\lambda)}$
$$
{\rm -\kappa c_{\kappa}(\eta;\lambda)} = c{\rm u_{g}^2(\eta;\lambda) +
\frac{d u_{g}(\eta;\lambda)}{d\eta}}
= {\rm -\kappa c - \frac{2}{c}
D^2_{\eta} \Big[ ln({\rm  I}_{\kappa}(\eta) + \lambda)}
\Big]
$$
\begin{equation} \label{-kc3}
= {\rm -\kappa c - \frac{4 w_{\kappa}(\eta) w_{\kappa}^{\prime}
(\eta)}{c({\rm  I}_{\kappa}(\eta)
+ \lambda)}
+ \frac{2 w_{\kappa}^4(\eta)}{c({\rm  I}_{\kappa}(\eta) + \lambda)^2}~,}
\end{equation}
where ${\rm {\rm I}_{\kappa}(\eta)= \int _{0}^{\eta} \,
w_{\kappa}^2(y)\, dy}$,
if we think of a half line problem for which $\lambda$ is a positive
integration constant thereby considered as a free parameter of the method..

All ${\rm  c_{\kappa}(\eta;\lambda)}$ have the same
supersymmetric partner index
${\rm c_{\kappa,susy}(\eta)
}$ obtained by deleting the zero mode
solution ${\rm w_{\kappa}}$.
They may be considered
as intermediates between the initial constant index ${\rm \kappa c}$ and
the supersymmetric partner
index ${\rm c_{\kappa,susy}(\eta)
}$.
From Eq.~(\ref{wbg}) one can infer the new parametric `zero mode' solutions of
the universe for the family of barotropic
indices ${\rm c_{\kappa}(\eta;\lambda)}$ as follows
\begin{equation} \label{wg}
{\rm w_g(\eta;\lambda)= 
\frac{w_{\kappa}(\eta)}{{\rm I}_{\kappa}(\eta) + \lambda}
\Longrightarrow a_g(\eta,\lambda)=
\left(\frac{w_{\kappa}(\eta)}{{\rm I}_{\kappa}(\eta) + \lambda}
\right)^{\frac{1}{c}}}~.
\end{equation}

Before closing this section, we recall an interesting point.
Since what we have done here is to use the Darboux transformations at the level of cosmological
evolutionary equations (i.e., equations of motion of the scale factor of the FRW cosmologies) 
a natural question is what is the effect of such transformations at the level of any Lagrangian of 
the cosmological fluid mechanics. The answer to this question has been already provided in 
the literature. Neto and Filho \cite{nf} have shown that in general the application of the Darboux transformations is equivalent 
to the addition of a total time derivative of a purely imaginary function to the Lagrangian and later, 
Samsonov \cite{sam} using the coherent state approach confirmed their result. 

\bigskip
{\bf 3} -  {\bf Dirac-like formalism}

\noindent
The Dirac equation in the susy nonrelativistic formalism has been discussed by Cooper {\em et al} \cite{cooper} already in 1988.
They showed that the Dirac equation with a Lorentz scalar potential is associated with a susy pair of Schroedinger Hamiltonians.
This result has been used later by many other authors. Here we make an application to barotropic FRW cosmologies that we find 
not to be a trivial exercise except for the uncoupled `zero-mass' case (subsection {\bf 3.1}). 

{\bf 3.1}-
Let's introduce now the following two Pauli matrices $\alpha =-{\rm i}\sigma _y=-{\rm i}\left( \begin{array}{cc}
0 & -{\rm i }\\
{\rm i} & 0\end{array} \right ) $ and $\beta =\sigma _x=\left( \begin{array}{cc}
0 & 1\\
1 & 0 \end{array} \right ) $ and write a cosmological Dirac equation
\begin{equation} \label{HD}
{\rm H_{D}^{{\rm FRW}}W=[i\sigma _y P_{\eta}+\sigma _x (icu_p)]W=0}~,
\end{equation}
where $W=\left( {\rm \begin{array}{cc}
w_1\\
w_2\end{array}} \right ) $ is a two component `zero-mass' spinor. 
This is equivalent to the following decoupled equations
\begin{eqnarray}
{\rm -P_{\eta}w_1+icu_pw_1=0}\\
+{\rm P_{\eta}w _2+icu_pw_2=0}~.
\end{eqnarray}
Solving these equations one gets ${\rm w}_1\propto 1/\cos ({\rm c}\eta)$ and ${\rm w}_2\propto \cos({\rm c}\eta)$ for $\kappa =1$ cosmologies
and ${\rm w}_1\propto 1/{\rm sinh} ({\rm c}\eta)$ and ${\rm w}_2\propto {\rm sinh}({\rm c}\eta)$ for $\kappa =-1$ cosmologies.
Thus, we obtain 
$$
W=\left( {\rm \begin{array}{cc}
w_1\\
w_2\end{array}} \right )=\left( {\rm \begin{array}{cc}
{\rm w_f}\\
{\rm w_b}\end{array}} \right )~.
$$
This shows that the matrix `zero-mass' Dirac equation is equivalent to the two linear second-order differential  
equations for the bosonic and fermionic components.  

\bigskip

{\bf 3.2}-
Consider now a ``massive" Dirac equation 
\begin{equation} \label{HDM}
{\rm H_{D}^{FRW}W=[i\sigma _y P_{\eta}+\sigma _x (icu_p +K)]W=KW}~,
\end{equation}
where $K$ may be considered the mass parameter of the Dirac spinor. Eq.~(\ref{HDM}) 
is equivalent to the following system of coupled equations
\begin{eqnarray}
{\rm -P_{\eta}w _1+(icu_p+K)w _1=Kw _2}\\
{\rm P_{\eta}w _2+(icu_p+K)w _2=Kw _1}~.
\end{eqnarray}
These two coupled first-order equations are equivalent to the following second order equations for the two spinor components
\begin{equation} \label{sch1der}
{\rm -P_{\eta}^{2}w _{i}-c\Big[ i(\mp P_{\eta}-2K)u_p+
 cu_{p}^{2}\Big]w _{i}=0}~,
\end{equation}
where the subindex $i=1,2$.

The fermionic spinor component can be found directly as solutions of
\begin{equation} \label{comp1} 
{\rm D^{2}_{\eta}w_1^{+}-\Big[c^2(1+2\tan ^2 c\eta)+2icK\tan c\eta\Big] w_1^{+}=0  \qquad {\rm for} \, \kappa =1}
\end{equation}
and
\begin{equation} \label{comp1b} 
{\rm D^{2}_{\eta}w_1^{-}-\Big[c^2(-1+2{\rm coth}^2 c \eta)-2icK{\rm coth} \,c\eta\Big] w_1^{-}=0  \qquad  {\rm for} \, \kappa =-1}~,
\end{equation}
whereas the bosonic components are solutions of
\begin{equation} \label{comp2} 
{\rm D^{2}_{\eta}w_2^{+}+\Big[c^2-2icK\tan c\eta\Big] w_2^{+}=0  \qquad {\rm for} \quad  \kappa =1}
\end{equation}
and 
\begin{equation} \label{comp2b} 
{\rm D^{2}_{\eta}w_2^{-}+\Big[-c^2+2icK{\rm coth} \,c\eta\Big] w_2^{-}=0  \qquad {\rm for} \quad \kappa =-1}~.
\end{equation}
The solutions of the bosonic equations are expressed in terms of the Gauss hypergeometric functions $_2F_1$ in the variables ${\rm y=e^{ic\eta}}$ and
${\rm y=e^{c\eta}}$, respectively
$$
{\rm w_2^{+}=Ay^{-p}\, _{2}F_1\Big[-\frac{1}{2}(p+iq);-\frac{1}{2}(p-iq), 1-p; -y^2\Big]}+
$$
\begin{equation} \label{s1}
{\rm By^{p} \, _2F_1\Big[\frac{1}{2}(p-iq), \frac{1}{2}(p+iq),1+p; -y^2\Big]}
\end{equation}
and 
$$
{\rm w_2^{-}=C(-1)^{-\frac{i}{2}r}y^{-ir}\, _{2}F_1\Big[-\frac{i}{2}(r+is),-\frac{i}{2}(r-is), 1-ir; y^2\Big]}+ 
$$
\begin{equation} \label{s2}
{\rm D(-1)^{\frac{i}{2}r}y^{ir}\, _{2}F_1\Big[\frac{i}{2}(r-s),\frac{i}{2}(r+s), 1+ir; y^2\Big]}~,
\end{equation}
respectively. The parameters are the following: ${\rm p=(-1-\frac{2K}{c})^{\frac{1}{2}}}$, ${\rm q=(1-\frac{2K}{c})^{\frac{1}{2}}}$,
${\rm r=(-1-i\frac{2K}{c})^{\frac{1}{2}}}$, ${\rm s=(-1+i\frac{2K}{c})^{\frac{1}{2}}}$, whereas $\rm A$, $\rm B$, $\rm C$, $\rm D$ are superposition 
constants.

It is not necessary to try to find the general fermionic solutions through the analysis of their differential equations (\ref{comp1}) and (\ref{comp1b}) 
because they are related in a known way to the bosonic solutions.\cite{ros}
The general fermionic solutions can be obtained easily if one argues that the particular fermionic zero mode is the inverse of a particular
bosonic zero mode and constructing the other independent zero mode solution as in textbooks. Thus
\begin{equation}\label{s3}
{\rm w_1}^{\pm}=\frac{1 +k\int ^{y}[{\rm w_2}^{\pm}]^2dz}{{\rm w_2}^{\pm}}~,
\end{equation}
where $k$ is an arbitrary constant.

\bigskip

{\bf 3.3}-
The most general case in this scheme is to consider the following matrix Dirac-like equation
$$
\Bigg[{\rm i}\left( \begin{array}{cc}
0 & -{\rm i }\\
{\rm i} & 0\end{array} \right ){\rm P_{\eta}}+\left( \begin{array}{cc}
0 & 1\\
1 & 0 \end{array} \right )\left( \begin{array}{cc}
 {\rm icu_p +K_1}& 0\\
0 &{\rm  icu_g+K_2}\end{array} \right )\Bigg]\left( \begin{array}{cc}
{\rm w}_1\\
{\rm w}_2 \end{array} \right )=
$$
\begin{equation} \label{Dg}
\left( \begin{array}{cc}
{\rm K_1}& 0\\
0 &{\rm  K_2}\end{array} \right )\left( \begin{array}{cc}
{\rm w_1}\\
{\rm w_2} \end{array} \right )~.
\end{equation}
Proceeding as in {\bf 3.2} one finds the coupled system of first-order differential equations
\begin{eqnarray}
\Big[{\rm P_{\eta}}+{\rm icu_g}+{\rm K_2}\Big]{\rm w_2}={\rm K_1}{\rm w_1}\\
\Big[-{\rm P_{\eta}}+{\rm ic u_p}+{\rm K_1}\Big]{\rm w_1}={\rm K_2}{\rm w_2}
\end{eqnarray}
and the equivalent second-order differential equations
$$
-{\rm P_{\eta}}^{2}{\rm w} _{i}+\Big[{\rm ic(u_p-u_g)}+({\rm K_1-K_2})\Big]{\rm P_{\eta}}{\rm w}_i+
$$
\begin{equation} \label{Schrg}
\Big[ {\rm ic}(\pm {\rm P_{\eta}}{\rm u}_i+{\rm K_1}{\rm u_g}+{\rm K_2}{\rm u_p})
 -{\rm c^2}{\rm u_p u_g}\Big]{\rm w} _{i}=0~,
\end{equation}
where the subindex $i=1,2$, and ${\rm u_1}$ and ${\rm u_2}$ correspond to ${\rm u_p}$ and ${\rm u_g}$, respectively.
In the ${\rm D_{\eta}}$ notation this equation reads
$$
{\rm D_{\eta}}^{2}{\rm w} _{i}+\Big[{\rm c}\Delta {\rm u_{pg}}-{\rm i}\Delta {\rm K})\Big]{\rm D_{\eta}}{\rm w}_i+
$$
\begin{equation} \label{Schrgb}
+\Big[{\rm c}(\pm {\rm D_{\eta}}{\rm u}_i+({\rm iK}_1{\rm u_g}+{\rm K_2}{\rm u_p}))
 -{\rm c}^2{\rm u_{p}u_g}\Big]{\rm w} _{i}=0~.
\end{equation}
Under the gauge transformation 
\begin{equation} \label{gauge}
{\rm w} _{i}=z_{i}\exp \left(-\frac{1}{2}\int ^{\eta}\Big[{\rm c}\Delta {\rm u_{pg}}-{\rm i}\Delta {\rm K})\Big]d\tau\right)=z_i(\eta)\frac{e^{\frac{1}{2}{\rm i}\eta \Delta {\rm K}}}{({\rm I_{\kappa}+\lambda})^{\frac{1}{2}}}\end{equation}
one gets
\begin{equation} \label{schz}
-{\rm P_{\eta}}^{2}z_i+Q_i(\eta)z_i=0,  \quad {\rm or} \quad {\rm D_{\eta}}^{2}z_i+Q_i(\eta)z_i=0,
\end{equation}
where 
$$
Q_i(\eta)=\Big[{\rm c}(\pm {\rm D_{\eta}u}_i+({\rm iK_1}{\rm u_g}+{\rm K_2}{\rm u_p}))
 -{\rm c}^2{\rm u_{p}u_g}\Big]-
$$
\begin{equation} \label{Q}
\frac{1}{2}{\rm D_{\eta}}\Big[{\rm c}\Delta {\rm u_{pg}}\Big]-\frac{1}{4}\Big[{\rm c}\Delta {\rm u_{pg}}-{\rm i}\Delta {\rm K})\Big]^2
\end{equation}
for $i=1,2$, respectively. $Q_i$ are complicated `potential' functions and we were not able to find analytical solutions of Eq.~(\ref{schz}).

The corresponding Dirac spinor is of the following form
$$
W(\lambda,{\rm K_1,K_2})=\left( {\rm \begin{array}{cc}
z_1({\rm K_1})\\
z_2(\lambda ,{\rm  K_2})\end{array}} \right )=\left( {\rm \begin{array}{cc}
 {\rm w}_{\rm f}({\rm K_1})\\
{\rm w}_{\rm g}(\lambda, {\rm K_2})\end{array}} \right )~,
$$
where $ {\rm w}_{\rm g}(\lambda , {\rm K_2})$ is given by Eq.~(\ref{wg}) for ${\rm K_1=K_2=0}$.
For $\lambda \rightarrow \infty$ one obtains $W(\lambda, 0,0) \rightarrow W$. In addition, ${\rm u_g \rightarrow u_p}$ and for ${\rm K_1=K_2=K}$ one gets the 
particular case in {\bf 3.2}.

\bigskip
{\bf 4 - Conclusions}

\noindent
We come now to the interpretation of the mathematical results that we displayed in 
the previous sections. An examination of the formulas (23-26) and (37) show that the 
parameters $K$ introduce an imaginary part in the adiabatic index of the cosmological 
fluid. Thus, the supersymmetric techniques presented in this research letter are a particular 
way to consider dissipation and instabilities in the ideal case of barotropic FRW cosmologies.
More general scale factors of barotropic FRW universes incorporating a well-defined
type of dissipation can be obtained from the `zero-modes'
${\rm w_{1,2}}^{\pm}$ by means of the relation ${\rm a} \sim {\rm w}^{1/{\rm c(\eta;K)}}$. The indices 
${\rm c(\eta;K)}$ are redefined adiabatic indices that can be infered from the formulas 
(23-26) and (37), respectively.

\bigskip

\end{document}